# Do Time Delay and Investment Decisions: Evidence from an Experiment in Tanzania[☆]

Plamen Nikolov[abcd]

**Abstract.** Attitudes toward risk underlie virtually every important economic decision an individual makes. In this experimental study, I examine how introducing a time delay into the execution of an investment plan influences individuals' risk preferences. The field experiment proceeded in three stages: a decision stage, an execution stage and a payout stage. At the outset, in the *Decision Stage* (*Stage 1*), each subject was asked to make an investment plan by splitting a monetary investment amount between a risky asset and a safe asset. Subjects were informed that the investment plans they made in the Decision Stage are binding and will be executed during the *Execution Stage* (*Stage 2*). *The Payout Stage* (*Stage 3*) was the payout date. The timing of the *Decision Stage* and *Payout Stage* was the same for each subject, but the timing of the Execution Stage varied experimentally. I find that individuals who were assigned to execute their investment plans later (i.e., for whom there was a greater delay prior to the *Execution Stage*) invested a greater amount in the risky asset during the *Decision Stage* (*JEL* D03, D81, D91, OI10, O12, O16)

[☆]I thank the editor and two anonymous reviewers for comments that considerably improved the paper. Plamen Nikolov gratefully acknowledges research support by The Harvard Institute for Quantitative Social Science, the Economics Department at the State University of New York (Binghamton), the Research Foundation for SUNY at Binghamton. I thank Matthew Bonci and Matthew Slater for outstanding research support. This paper was previously circulated as a working paper entitled "Are Risk Preferences Time Inconsistent? Evidence from An Investment Experiment." All remaining errors are my own.

Corresponding Author: Plamen Nikolov, Department of Economics, State University of New York (Binghamton), Department of Economics, 4400 Vestal Parkway East, Binghamton, NY 13902, USA. Email: pnikolov@post.harvard.edu

[a] State University of New York (Binghamton)
[b] IZA Institute of Labor Economics
[c] Harvard Institute for Quantitative Social Science
[d] Global Labor Organization

# 1. Introduction

Attitudes toward risk underlie virtually every important economic decision an individual makes. They influence whether one engages in negative health behaviors (Anderson and Mellor 2008), what level of education to pursue (Belzil and Leonardi 2007), what type of employment to seek (Sapienza et al. 2009), when to marry (Spivey 2010), and how many children to have (Rodriguez and Seigle 2012). In addition, previous studies have found that individuals who exhibit higher levels of risk aversion, that is, the tendency to avoid better-than-fair gambles, also tend to exhibit higher levels of impatience or time discounting, which is the rate at which future costs and benefits are adjusted to make them comparable with current ones (Weber and Chapman 2005). In this study, I use data from a simple randomized artefactual intervention[1] conducted in Tanzania to examine how introducing a time delay into the execution of an investment plan influences individuals' risk preferences.

We recruited a sample of 350 participants in Morogoro, Tanzania for an experiment that proceeded in three stages. At the outset, in the *Decision Stage* (*Stage 1*), each subject was asked to make an investment plan by splitting a monetary investment amount (*I*) between a risky asset (*R*) and a safe asset (*I-R*). The risky asset had a 50 percent chance of returning five times the amount invested (*5R*) and a 50 percent chance of returning nothing. The safe asset only returned the amount invested (*I-R*). Subjects were informed that the investment plans they made in the *Decision Stage* could not be changed during the *Execution Stage* (*Stage 2*). During the *Decision Stage*, subjects were informed that they would receive the money (*I*) to execute their *Decision Stage* plan during the *Execution Stage*, and were told when the *Execution Stage* would occur (i.e., the duration of the delay). The study team provided the money that was used during the *Execution Stage*, but reclaimed it upon completion of the *Execution Stage*. We informed individuals that the money each person placed in the "savings" cup (safe asset) would be stored there until the last session of the game, at which time they would recoup all the money they had deposited in the savings cup. Finally, the *Payout Stage* (*Stage 3*) was the payout date. Subjects returned and were informed, based on a flip of a coin, whether their risky asset was successful or not. The timing of the *Decision Stage* and *Payout Stage* was the same for each subject, but the timing of the *Execution Stage* varied experimentally. I then examined how the timing of the *Execution Stage* influenced individuals' allocation of their initial funds to risky assets (*R*) during the *Decision Stage*.

I find that individuals who were assigned to execute their investment plans later (i.e., for whom there was a greater delay prior to the *Execution Stage*) invested a greater amount in the risky asset during the *Decision Stage*. On average, subjects across all treatment groups allocated 67 percent of their initial endowment to the risky asset and 33 percent to the safe asset. Each one-week delay in the *Execution Stage* increased the amount allocated to the risky asset in the *Decision Stage* by, on average, 57 Tanzanian shillings (TZS) (approximately 0.05 USD), which was approximately 3 percent of the initial endowment. Cumulatively, a four-week delay in the execution of an investment plan resulted in approximately 11 percent more of the endowment being invested in the risky asset.

After establishing that a time delay has a positive influence on one's willingness to invest in a risky asset, I examined the theoretical explanations of this empirical result. Although inconsistent with the results of a standard economic model on intertemporal decision-making, this result is consistent with a *reference-dependent* framework that incorporates a reference point (affected via temporal priming) within an individual's intertemporal decision-making process. I then developed a simple stylized model that explains how an increase in the lag between the time at which an investment plan is made and the time when that investment plan is executed increases the amount an individual invests in a risky asset rather than a safe asset.

---

[1] Based on the taxonomy presented in Harrison and List (2004).

Although recent studies have examined violations of expected utility theory[2] (e.g., when rational individuals respond to a cognitive bias that makes them choose differently depending on whether a prospect is presented as a loss or a gain), the studies that are most closely related to the current study are those of Noussair and Wu (2006) and Shelley (1994), who examine the stability of risk preferences and individual discount rates, respectively. Noussair and Wu (2006) explore the stability of risk preferences between lotteries that are resolved and paid immediately versus those that are resolved and paid in the future and find that subjects exhibit a greater level of risk aversion toward lotteries that are resolved and paid immediately. Unlike the current study's setup, Noussair and Wu (2006) vary both the lottery choice and the payout date. Shelley (1994) examines how individuals discount risky lotteries and finds that depending on whether the payoffs are in terms of gains or losses, lotteries consisting exclusively of losses are discounted more heavily.[3]

Although classical economic models posit that individual preferences are stable across time and contexts, recent studies on endogenous preferences have challenged this assumption and argued that contextual factors, such as priming, can shape individual preferences. Priming, a concept that was first developed in the field of psychology (Meyer and Schvaneveldt 1971, Neeley 1977, Tulving and and Murray 1985, Dehaene et al. 2006, Bargh and Chartrand 2000, Doyen et al. 2012), refers to the activation of mental concepts through subtle situational cues or stimuli that can influence one's judgment and behavior as a result of prior encounters with the same or related stimuli.[4] Priming has recently also become the focus of economic studies, especially in the context of how situational cues influence one's economic decision-making.[5] For example, Lichand and Mani (2016) examine how exogenous shocks, in the form of weather shocks, can decrease an individual's attention, memory, and impulse control and increase their susceptibility to a variety of behavioral biases.[6] In the study setup presented here, the information regarding the timing of the *Execution Stage* serves as a form of temporal priming.[7] This information likely increases the salience of the present in a subject's mind.

---

[2] A related set of recent studies focuses on testing expected utility theory more broadly, in particular on the independence axiom in intertemporal dimensions (the Allais paradox and the common ratio effect). Baucells and Heukamp (2009), Gneezy, List and Wu (2006), and Andreoni and Sprenger (2010) examine behavior related to the common ratio effect. Gneezy, List, and Wu (2006) document a violation of this condition in which individuals value a risky prospect less than its worst possible realization. Andreoni and Sprenger (2010) show that subjects violate predictions of the common ratio effect in intertemporal contexts.
[3] Within the economics literature, Anderhub et al. (2001), Anderson and Stafford (2009) and Epper, Fehr-Duda, and Bruhin (2009) explore various channels for how people's risk-taking behavior is interlinked with their time discounting.
[4] Based on the type, priming can occur via the speed of processing (Reisberg 2007), via items of a similar form or meaning (Biederman and Cooper 1992), via exposure repetition (Forster and Davis 1984), via items sharing similar semantic features (Ferrand and New 2003), or via stimuli that trigger visuomotor system effects (Klotz and Wolff 1995). In addition, priming has been shown to influence subsequent behavior in the context of various stimuli: visual, spatial, physical, olfactory, and verbal cues (Kay et al. 2004, Biederman and Cooper 1992), and most recently temporal stimuli (Trope and Liberman 2000, Huber et al. 2002, Blandin and Dehaene 2002, Fujita et al. 2006, Kivetz and Tyler 2007, Ebert and Prelec 2007, Mannetti et al. 2009, Naccache, Zauberman et al. 2009, Yashar and Lami 2010, Bauer, Muller and Usher 2009).
[5] For recent economic studies that explicitly consider priming in individual decision-making, see Matthey (2010), Benjamin, Choi, and Strickland (2010), and Cohn et al. (2015).
[6] Shah, Shafir, and Mullainathan (2015) document various recent behavioral biases, including priming, in the context of economic decision-making.

This study makes two important contributions to the economics literature. First, it provides evidence that temporal priming can influence individual behavior in the context of important economic outcomes (e.g., investment decisions) by showing that it can induce individuals to choose a more risky option. Knowing which factors can induce individuals to take on more risk can be useful in various policy contexts. Previous economic studies have largely focused on conceptual priming and its influence on behavior and how social identity shapes one's preferences (Cohn and Marechal 2016), whereas this study focuses on whether temporal priming can influence one's behavior.[8] Second, this study provides evidence of time inconsistency in the domain of investment behavior. This differs from recent experimental studies on individual time inconsistency that have largely examined behavior in the consumption domain (e.g., Ashraf, Karlan and Yin 2006).[9] In contrast, this paper provides evidence of time inconsistency in the context of investment decision-making.

The rest of the paper proceeds as follows. Section 2 presents a simple theoretical model and testable predictions. Section 3 describes the experimental design, data, and study sample. Section 4 presents the results. Section 5 presents a theoretical explanation of the empirical findings and Section 6 concludes.

## 2. A Simple Framework with a Delay in Investment Execution

I outline a simple, classical decision-making framework to highlight a key prediction regarding this study's experimental design. Let us assume that an individual faces a timeline with three time points: $t_0$ (investment plan creation), $t_1$ (pre-made investment plan execution), and $t_2$ (investment payout). At $t_0$, the individual makes a plan involving the spread of investment of a fixed monetary endowment between two assets: a risky asset and a safe asset. There are two states of the world to consider: good (G) and not good (NG). An individual chooses between two gambles, A and B, where:

$$A = (p_G^A, p_{NG}^A, c_G^A, c_{NG}^A) \text{ and } B = (p_G^B, p_{NG}^B, c_G^B, c_{NG}^B)$$

where $p_G^i$ is the probability of state G, $p_{NG}^i$ is the probability of state NG, $c_G^i$ is the payout in state G, $c_{NG}^i$ is the payout in state NG, and i = A, B. These gambles have the following expected utilities:

$$EU^A = p_G^A \times c_G^A + p_{NG}^A \times c_{NG}^A \text{ and } EU^B = p_G^B \times c_G^B + p_{NG}^B \times c_{NG}^B$$

A is associated with payouts $(c_G^A, c_{NG}^A)$ and B is associated with payouts $(c_G^B, c_{NG}^B)$, with all payouts occurring at $t_2$. A time-consistent individual, whose preferences satisfy the completeness, transitivity, continuity, and independence axioms, will make a choice at $t_0$ by comparing the expected utilities associated with the two gambles. He or she will choose A over B if and only if, $EU^A > EU^B$. From a neo-classical standpoint, the timing of $t_1$ does not factor into the comparison between the two gambles. In other words, the amount invested in the riskier gamble should not vary with the timing of $t_1$. In a standard intertemporal model, an individual chooses between A and B by comparing $EU^A$ with $EU^B$ discounted from the perspective of time point $t_0$. For a time-consistent individual, *ceteris paribus*, the timing of $t_1$ (at which point one makes an investment based on a pre-existing and binding investment plan) does not influence one's decision, made at time point $t_0$, to split an amount of money $I$ between various gambles. This study sets up an experimental design to test this prediction.

---

[7] Several recent studies rely on cues related to temporal stimuli. See, e.g., Trope and Liberman (2000), Fujita et al. (2006), Kivetz and Tyler (2007), Ebert and Prelec (2007), Zauberman et al. (2009), Yashar and Lami (2010), and Bauer, Muller, and Usher (2009).
[8] Bargh and Chartrand (2000) and Jin (2015, pp. 16–60) discuss psychology studies that examine the role of various forms of priming.
[9] I also provide evidence of one way to influence reference point creation: visceral possession versus expectations (i.e., the Köszegi and Rabin model).

## 3. Experimental Design

### 3.1 Experimental Setup

We recruited 350 subjects, all of whom were over 18 years of age, from catchment areas with the help of village and district leaders in Morogoro, Tanzania. We directly invited all individuals to participate in our study over a period of four weeks (see Table 1).

**Table 1:** Summary of Treatment Groups

|  | Treatment Group 1 | Treatment Group 2 | Treatment Group 3 | Treatment Group 4 | Treatment Group 5 |
|---|---|---|---|---|---|
|  | (1) | (2) | (3) | (4) | (5) |
| Endowment | 2,000 TZS | 2,000 TZS | 2,000 TZS | 2,000 TZS | 2,000 TZS |
| Sample Size | 70 | 70 | 70 | 70 | 70 |
| Delay (in weeks) to Execution Stage ($t_1$) | 0 | 1 | 2 | 3 | 4 |

*Notes*: At the time of the study, 1 U.S. dollar was approximately equivalent to 1,570 Tanzanian schillings (TZS).

Each individual in our three-stage study faced two investment choices: one was a risky asset (A), offering a high mean and variance, while the other was a safe asset (B). A had a 50 percent chance of returning five times the amount invested ($R$) and a 50 percent chance of returning nothing. B paid out the amount invested in it (i.e., the difference between the original money endowment ($I$) and $R$, or $I-R$). There were five groups of investors, and each group was assigned a different time for the execution of their investment plans (see Figure 1).

- **The Decision Stage** ($t_0$): During this stage, each subject made a plan regarding the investment of 2,000 Tanzanian shillings (TZS) (approximately 1.3 USD). The investment was to be spread between *A* and *B*. The subjects knew how long they would have to wait until execution (i.e., they were advised when the execution would occur) when they made their investment plans.
- **The Execution Stage** ($t_1$): During this stage, individuals were given the cash that was promised in the *Decision Stage* ($I$) and told to invest it in accordance with the investment plan made in the *Decision Stage*.[10]
- **The Payout Stage** ($t_2$): During this stage, the outcomes of lotteries chosen at $t_0$ were revealed (based on a flip of a coin, the risky asset was deemed either successful or otherwise), and each individual received his or her payout.

---

[10] The money used during the *Execution Stage* was reclaimed upon completion of the *Execution Stage* to avoid the possibility of an endowment effect. We informed individuals that the money each person placed in the "savings" cup (safe asset) would be stored safely there until the last session of the game, at which point they would receive all of the money deposited in the "savings" cup.

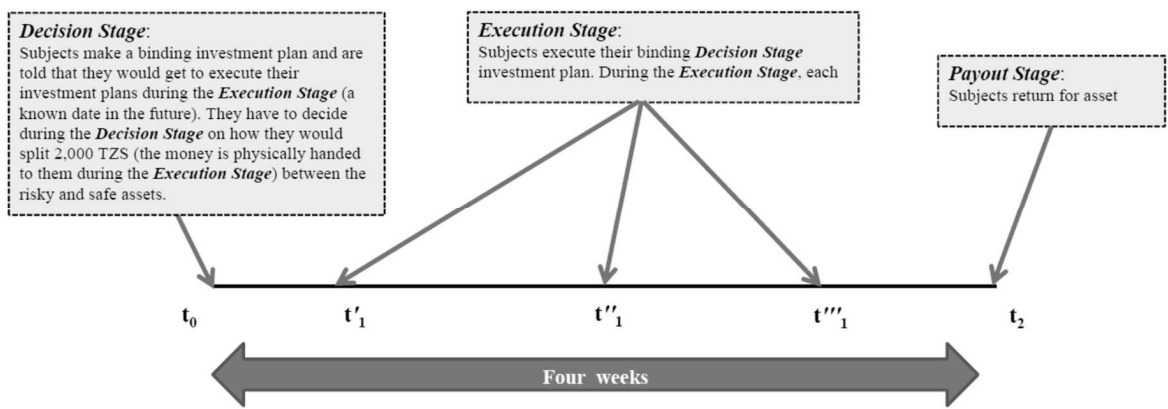

Figure 1—Stylized Timeline of the Study Design

The main objective of this experimental design was to randomly assign the timing of the *Execution Stage*. We randomly varied the timing of the *Execution Stage* to occur at weekly intervals between one and four weeks after the *Decision Stage*, and examined how, *ceteris paribus*, the portion of *I* that was invested in the risky asset A varied with the timing of the *Execution Stage*.

Each participant was given the following basic instructions in Swahili (see Appendix B for the detailed instructions):

> This game consists of three sessions. The first session will take place today. The second session will take place in this same venue on ____. The third session will occur in four weeks. It is very important that you attend all three sessions. We will give you a *letter* reminding you of the dates of the second and third sessions. In the first session of the game that will take place today, we will ask you to make an investment plan for a sum of money we will give you in <xx> weeks' time during the second session. You will have to decide today how to invest this money in <xx> weeks' time. At that time, you will divide the money between two account cups, a "savings" cup and a "business" cup, based on the investment plan you make today. The money that you decide to place in the business cup in <xx> weeks' time can generate a greater return than money placed into the savings cup; however, there is the possibility of no gain. The money that you decide today to place in the saving" cup in <xx> weeks as well as the money that you place in the business cup will be stored and given to you at the end of the third session. Now, we will explain the step-by-step details of the game.

As can be seen, during the *Decision Stage*, subjects were alerted to the specific timing of the *Execution Stage*. As discussed earlier, it is entirely possible that the specification of the timing of the *Execution Stage* increased its salience, and that some subjects were primed to think of short timeframes. During the *Decision Stage* and the *Execution Stage*, we used the term "cup" to refer to each hypothetical asset (i.e., a "business" cup and a "savings" cup).[11]

## 3.2 Data and Sample Characteristics

We used two sources of data. The first source was our background survey, which included questions on the baseline characteristics of the participants, such as marital status, household composition, assets, and health (the descriptive statistics are presented in Table 2). The second source of data elicited time and risk preferences from respondents (see Figure A2) and measured cognitive ability. The time-preference questions asked respondents to decide between receiving TZS 5,450 (approximately 3 USD) now and a larger amount a month later. To measure time consistency, we also asked respondents to choose between receiving TZS 5,450 in one month's time or a larger amount in two months' time. The risk-preference questions were similar to those of Charness and Genicot (2009), and asked respondents how much of an initial amount of TZS 1,800 (approximately 1.2 USD) they would like to invest in an asset that would return either four times the amount invested or nothing with probability 0.5 (see Figure A1). To measure cognitive ability, we asked respondents to complete a Raven's Progressive Matrices test, in which they had to recognize patterns in a series of images. This cognitive assessment generated a score between zero and eight.

---

[11] The money an individual decided to place in the business cup could be used to generate more money—like a business. During the *Execution Stage*, the study team visited the subjects, reminded them of the investment plans they had made during the *Decision Stage*, and confirmed that the plans were binding. Subjects were also reminded that investments placed in the business cup could either succeed or fail. Therefore, if the "business" succeeded, they would earn a return on the amount they had placed in the business cup. However, if the "business" failed, they would receive nothing from the "business" venture. After being provided with this information, subjects were asked to execute their *Decision Stage* plans by physically depositing the money that was handed to them into the appropriate cup or cups.

The study sample came from the Morogoro area, and no study participants were part of any other study at the time. Although we did not collect information on occupation, the majority of the study participants were engaged in the informal agricultural sector. The summary statistics in Table 2 show that the sample was divided approximately evenly between males and females, the average age of study participants was 36 years, approximately 18 percent of the study sample had completed primary education or above, and approximately half were married at the time of the study. Study participants had earned an average of 157,296 TZS (approximately 100 US dollars) in the month prior to the study and the mean number of children in their households was slightly more than two.

**Table 2:** Summary Statistics of Sample Subjects

|  | Mean (SD) |
|---|---|
| *Demographics* |  |
| Age | 36.44 (13.61) |
| Male (1=male, 0=female) | 0.48 (0.50) |
| Completed more than primary (yes=1, no=0) | 0.18 (0.38) |
| Currently married (1=yes, 0=no) | 0.49 (0.50) |
| Number of children in household | 2.32 (2.08) |
| Earnings last month (in TZS) | 157,296.43 (250,730.11) |
| Number of chickens (Assets) | 4.85 (10.99) |
| Metal roof (1=yes, 0=no) | 0.80 (0.39) |
| *Risk and Time Preferences* |  |
| Amount invested (out of 1800 TZS) in risky asset | 1,260.07 (430.60) |
| Somewhat patient | 0.15 (0.36) |
| Time-consistent | 0.25 (0.43) |
| Present-biased | 0.37 (0.48) |
| More patient in future than in present | 0.18 (0.39) |
| *Cognitive Skills* |  |
| Ravens ability test (# correct out of eight questions) | 3.26 (2.08) |
| Observations | 350 |

*Notes*: Standard deviations presented in parentheses. At the time of the study, 1 U.S. dollar was approximately equivalent to 1,570 Tanzanian schillings (TZS). The risky asset paid off five times the amount invested with probability 0.5, and 0 with probability 0.5. Risk Preference Module is in Appendix Figures. "Somewhat Patient" is a dummy variable equal to 1 if the respondent prefers TZS 5,650 (or less) in a month to TZS 5,450 now (Time Module is in Appendix Figures), and 0 otherwise. "Time-Consistent" is a dummy variable equal to 1 if the respondent exhibits the same discount rate between today and one month from today, and 0 otherwise. "Present-Biased" is a dummy variable equal to 1 if the respondent exhibits a higher discount rate between today and one month from today than between one month from today and two months from today, and 0 otherwise. "More Patient in Future than in Present" is a dummy variable equal to 1 if the respondent will be more patient in one month than they are today, and 0 otherwise.

Using data from the demographic variables, I verified that no correlation existed between individual characteristics and the five randomly assigned treatment groups (see Table 3). Using two statistical tests regarding equality of means across treatment groups, I found that none of the variables exhibited statistical significance. This pattern is consistent with the successful implementation of random assignment.

Of the initial study sample, 335 participants (95 percent) completed the experiment. There were two main sources of attrition. Some participants could not be found at the time of the *Execution Stage*, and several participants failed to appear at the *Payout Stage*. The results presented in Table A2 in the Appendix show that there was no statistical evidence that attrition status was correlated with treatment group.

**Table 3:** Average Subject Sample Characteristics

|  | Group 1 | Group 2 | Group 3 | Group 4 | Group 5 | p-value joint test 1 | p-value joint test 2 |
|---|---|---|---|---|---|---|---|
|  | (1) | (2) | (3) | (4) | (5) | (6) | (7) |
| Age | 36.20 (12.29) | 34.40 (13.00) | 36.34 (12.73) | 37.71 (15.01) | 37.74 (15.09) | 0.66 | 0.88 |
| Male (1=yes, 0=no) | 0.44 (0.50) | 0.55 (0.50) | 0.49 (0.50) | 0.52 (0.50) | 0.39 (0.49) | 0.27 | 0.49 |
| More than primary (1=yes, 0=no) | 0.13 (0.34) | 0.39 (0.49) | 0.19 (0.39) | 0.10 (0.30) | 0.11 (0.32) | 0.23 | 0.19 |
| Married (1=yes, 0=no) | 0.46 (0.50) | 0.41 (0.50) | 0.56 (0.50) | 0.58 (0.49) | 0.49 (0.50) | 0.23 | 0.42 |
| Children in household (Number) | 2.63 (1.84) | 1.65 (1.91) | 2.31 (2.49) | 2.49 (2.49) | 2.57 (2.34) | 0.11 | 0.25 |
| Monthly earnings (in TZS) | 146,157.89 (195,079.82) | 139,373.08 (198,253.35) | 165,458.33 (292,477.46) | 125,547.62 (182,327.53) | 209,500.00 (342,224.70) | 0.57 | 0.76 |
| Chickens (Number) | 7.12 (14.28) | 5.48 (9.90) | 4.93 (12.98) | 4.73 (11.25) | 2.55 (5.39) | 0.33 | 0.13 |
| Metal roof (1=yes, 0=no) | 0.77 (0.43) | 0.83 (0.38) | 0.80 (0.41) | 0.85 (0.36) | 0.79 (0.41) | 0.80 | 0.44 |
| Risk aversion | 1244 (443.96) | 1331 (446.81) | 1327 (403.37) | 1222 (434.20) | 1160 (413.27) | 0.15 | 0.79 |
| Somewhat patient | 0.08 (0.28) | 0.17 (0.38) | 0.21 (0.41) | 0.16 (0.37) | 0.13 (0.13) | 0.39 | 0.13 |
| Time-consistent | 0.24 (0.43) | 0.20 (0.41) | 0.33 (0.47) | 0.29 (0.46) | 0.21 (0.41) | 0.49 | 0.85 |
| Present-biased | 0.31 (0.47) | 0.36 (0.48) | 0.34 (0.49) | 0.42 (0.49) | 0.44 (0.50) | 0.52 | 0.25 |
| More patient in future than in present | 0.18 (0.39) | 0.17 (0.38) | 0.20 (0.40) | 0.19 (0.39) | 0.16 (0.37) | 0.98 | 0.95 |
| Raven's test (number correct) | 2.87 (2.36) | 3.68 (1.84) | 3.37 (1.97) | 3.17 (2.03) | 3.15 (2.18) | 0.21 | 0.13 |
| Observations | 70 | 70 | 70 | 70 | 70 |  |  |

*Notes:* Standard deviations presented in parentheses. At the time of the study, 1 U.S. dollar was approximately equivalent to 1,570 Tanzanian schillings (TZS).
Joint test 1: Test of equality of means across four treatment groups (i.e., complete null hypothesis).
Joint test 2: Joint test that means in treatment groups (Groups 2-5) are equal to mean in control group (Group 1).

## 4. Results: Intent-to-Treat Analysis
### 4.1 Intent-to-Treat Effect: Impact on Investment in Risky Assets

The main outcome of interest, and thus our dependent variable, is the amount invested in the risky asset (Y). Let us assume that $Z_{G2}$ is an indicator variable for assignment to Treatment Group 2, which is allowed to execute its investment plan one week after the plan is made, $Z_{G3}$ is an indicator variable for assignment to Treatment Group 3, which is allowed to execute its investment plan two weeks after the plan is made, $Z_{G4}$ is an indicator variable for assignment to Treatment Group 4, which is allowed to execute its investment plan three weeks after the plan is made, and $Z_{G5}$ is an indicator variable for assignment to Treatment Group 5, which cannot execute its investment plan until four weeks after the plan is made. I estimate the following equation for the full sample:

$$Y_{it} = \beta_0 + \beta_1 X_1 + \beta_{G2} Z_{G2,i} + \beta_{G3} Z_{G3,i} + \beta_{G4} Z_{G4,i} + \beta_{G5} Z_{G5,i} + \varepsilon_I \quad (1)$$

where $X_i$ is a vector of additional controls and $\beta_{G2}$, $\beta_{G3}$, $\beta_{G4}$, and $\beta_{G5}$ are coefficients that capture the causal effect of treatment group assignment to Groups 2, 3, 4, and 5, respectively, relative to the control group (Treatment Group 1). Individuals in the control group (Treatment Group 1) execute their investment plan without delay. $\beta_{G2}$ provides an estimate for the intent-to-treat (ITT) effect—an average of the causal effects—for an individual who is assigned to execute his investment plan after a one-week delay and so on for $\beta_{G3}$ (a two-week delay), $\beta_{G4}$ (a three-week delay), and $\beta_{G5}$ (a four-week delay). Table 4 presents the results, clustered at the session level, for the experimental subjects' investment behavior by treatment group (the reference group is Treatment Group 1).

**Table 4:** Intent-to-Treat Analysis of Investment in Risky Asset

|                    | Amount in Risky Asset | Amount in Risky Asset |
|--------------------|-----------------------|-----------------------|
|                    | (1)                   | (2)                   |
| Treatment Group 2  | 127.88***             | 109.96***             |
|                    | (44.78)               | (31.51)               |
| Treatment Group 3  | 171.34***             | 147.65***             |
|                    | (38.34)               | (25.56)               |
| Treatment Group 4  | 214.60***             | 184.10***             |
|                    | (65.70)               | (62.56)               |
| Treatment Group 5  | 268.02***             | 225.98***             |
|                    | (40.50)               | (45.86)               |
|                    |                       |                       |
| Mean Dep Variable  | 1329                  | 1329                  |
| Covariates         | No                    | Yes                   |
| Observations       | 350                   | 350                   |
| $R^2$              | 0.11                  | 0.13                  |

*Notes*: All results are in reference to Group 1. Standard errors in clustered at the session level. At the time of the study, 1 U.S. dollar was approximately equivalent to 1,570 Tanzanian schillings (TZS). Controls include gender, age, marital status, number of children, asset proxies, risk aversion measurement, time preferences, Raven's test total score. ***, ** and * indicate significance at the 1, 5 and 10 percent levels.

Overall, Table 4 provides strong evidence that as the delay until execution increases, subjects display a greater willingness to invest more of their monetary endowment in the risky asset. *Ceteris paribus*, a one-week delay prior to execution of the investment plan results in an increase of 57 TZS (approximately 0.05 USD), or approximately 3 percent, of each subject's initial endowment in the amount invested in the risky asset. Thus, a four-week delay prior to execution of the investment plan results in approximately 11 percent more of the initial endowment being invested in the risky asset.

**4.2 Quantile Treatment Effects**

Estimation of the quantile treatment effects shows the distribution of impacts and avoids the possibility of drawing misleading conclusions based on outliers. Table A3 shows regressions, clustered at the session level, for quintiles of the distribution. The estimated treatment effect at the tenth percentile may be interpreted as the difference in balance changes between two individuals—one in the treatment group and the other in the control group (i.e., Group 1) —positioned at the tenth percentile of the distribution of changes within their respective groups. Comparing the treatment group with the control group, the largest treatment effects are at the top of the distribution.

**4.3 Heterogeneous Treatment Effects**

Next, I examined differential impacts in relation to several demographic and behavioral characteristics. Table 5 shows the results of the same regressions (clustered at the session level) shown in Table 3, but with the treatment indicator variable interacted with one demographic or behavioral variable at a time. The demographic and behavioral variables included gender, education, earnings, and cognitive skills. The coefficient for the interaction term is insignificant for all variables. This suggests that within the treatment group, the average effect of the treatment assignment is operating uniformly across these other characteristics.

**Table 5:** Intent-to-Treat Analysis of Subgroups. Investment in Risky Asset

|  | Amount in Risky Asset (1) | Amount in Risky Asset (2) | Amount in Risky Asset (3) | Amount in Risky Asset (4) |
|---|---|---|---|---|
| Treatment Group 5 | 189.99*** (78.29) | 247.71*** (48.08) | 154.58*** (60.84) | 122.55** (77.30) |
| Male | -30.03 (83.46) |  |  |  |
| Male × Treatment Group 5 | 28.93 (118.51) |  |  |  |
| Primary completed |  | 60.65 (71.58) |  |  |
| Primary completed × Treatment Group 5 |  | -215.85 (146.51) |  |  |
| Earnings |  |  | 0.00 (0.00) |  |
| Earnings × Treatment Group 5 |  |  | 0.00 (0.00) |  |
| Raven's cognitive score |  |  |  | 8.87 (7.67) |
| Raven's cognitive score × Treatment Group 5 |  |  |  | 21.01 (17.27) |
| Covariates | Yes | Yes | Yes | Yes |
| Observations | 150 | 150 | 150 | 150 |
| $R^2$ | 0.37 | 0.40 | 0.40 | 0.38 |

*Notes*: All results are in reference to Group 1. Standard errors in clustered at the session level. At the time of the study, 1 U.S. dollar was approximately equivalent to 1,570 Tanzanian schillings (TZS). Controls include gender, age, marital status, number of children, asset proxies, risk aversion measurement, time preferences, Raven's test total score out of eight questions. ***, ** and * indicate significance at the 1, 5 and 10 percent levels.

## 5. Discussion of Potential Mechanisms: A Stylized Model with Reference-Dependent Risk Preferences

I present a simple model wherein agents make investment decisions based not on the final outcome, but rather on the potential losses and gains from the investment. Each agent determines these potential values by evaluating prospective losses and gains using a certain heuristic process formed from their own perspective. At $t_0=0$, an individual decides whether to invest or save an endowment ($I=1$) that he or she will receive at $t_1$. The gains from saving and investment are realized at $t_2$ subject to the following specifications:

- *Investment*: earns $k$ with probability $(\frac{1}{2})$ and 0 with probability $(\frac{1}{2})$; and
- *Saving*: the savings rate is such that the net present value (NPV) of the endowment is unaffected.

To find the minimum level of k that induces the agent to invest, I specify a framework relating the individual's decision to $t_1$. The individual's value function features three characteristics. First, it features reference dependence (i.e., the carrier of an attribute's value is not based on its absolute level, but rather on its deviation from some reference level, which results in either a gain or a loss). Second, the value function is steeper for losses than for gains (i.e., a loss decreases value more than a gain of equivalent size increases value). Third, I embed diminishing sensitivity (i.e., the marginal value of both gains and losses decreases with their size; the first sip of beer tastes the best and the first dollar lost hurts the most). For simplicity, I formulate linear utility as follows:

$$U(w|\rho) = \begin{cases} w - \rho & if \quad w \geq \rho \\ \gamma(w - \rho) & if \quad w < \rho \end{cases} \quad (2)$$

with $\gamma > 1$ (typically $\gamma = 2$).[12] Thus, for reference structures that satisfy constant loss aversion, changes in the preference order induced by a shift in a reference point can be described in terms of the constant $\gamma$, which is the coefficient of loss aversion.

I assume that the reference point is the discounted endowment value $\rho = \delta^{t_1}$.[13] The intuition behind this assumption is that investing money that one has already received, or will soon receive, is more likely to bring an individual into the loss domain than investing money that will be received in the future. In other words, the further into the future the execution of a planned investment is, the more an individual is willing to demote the perception of a "prize" associated with carrying out the plan. Time point $t_1$ manipulates, very likely through temporal priming with the temporal reminder regarding the timing of $t_1$ (the *Execution Stage*), the individual's perception of derived gains and derived losses.

Substituting $\rho = \delta^{t_1}$ into (3), I get:

$$U(w|\delta, t_1) = \begin{cases} w - \delta^{t_1} & if \quad w \geq \delta^{t_1} \\ \gamma(w - \delta^{t_1}) & if \quad w < \delta^{t_1} \end{cases} \quad (3)$$

The lottery prospect offers a gain with probability 0.5 ($u = \delta^{t_2}k - \delta^{t_1}$) and a loss with probability 0.5 ($u = -\gamma\delta^{t_1}$). To maximize his or her expected utility, an individual chooses to invest if $\frac{1}{2}(k\delta^{t_2} - \delta^{t_1}) - \frac{1}{2}\gamma\delta^{t_1} > 0$. This expression provides us with $k$ (i.e., the investment amount) as a function of $t_1$ and $\delta$:

$$k > \frac{(1+\gamma)\delta^{t_1}}{\delta^{t_2}} \equiv k^* \quad (4)$$

Based on (4), our key prediction is that an increase in the lag between the planning and execution of an investment increases an individual's propensity to invest rather than save. The intuition behind this prediction is that the reference point for a loss or a gain is lower for delayed earnings, and this delay reduces loss aversion.

## 6. Conclusion

Using data from an artefactual experiment, I estimate the causal effects of introducing a delay prior to the execution of an investment plan on how individuals allocate monetary endowments between a safe and a risky asset. I find a significant positive effect of a time delay on the amount individuals are willing to invest in a risky asset, all else being equal. On average, subjects across all treatment groups allocated 67 percent of their initial endowment to the risky asset and 33 percent to the safe asset. A one-week delay in the execution of the investment plan induced subjects to allocate 3 percent more of their endowment to the risky asset when planning their investment allocation. Cumulatively, a four-week delay in the execution of an investment plan resulted in approximately 11 percent more of their endowment being invested in the risky asset.

---

[12] Empirical estimates of the loss aversion coefficient are approximately around 2 (Tversky and Kahneman 1992)
[13] This assumption is supported by Mohr et al. (2010), Tu (2004), Oxoby and Morrison (2010), McAlvanah (2007), and Baucells and Heukamp (2009), and is ingrained in the intertemporal version of the Weber–Fechner Law in Tversky and Kahneman (1992) and construal-level theory in Liberman, Sagristano, and Trope (2002).

      The findings of this study raise a number of issues concerning the pathways through which a time delay influences one's willingness to invest in riskier assets. Due to this study's design, I cannot rule out alternative explanations of the key result. Thus, future studies should focus on identifying the mechanisms driving this empirical result. The results of the study draw attention to an empirical finding that may have important implications for human behavior. This finding also highlights an area in which the predictions of standard economic models are wide of the mark. A detailed examination of how individuals respond to temporal priming cues in the intertemporal decision-making model could, in some settings, not only provide better predictions of behavior, but also be welfare-enhancing, with potential implications for policy-makers.